# Scaling behavior of columnar structure during physical vapor deposition

W. J. Meese[1], T.-M. Lu[2]

Department of Physics, Applied Physics, and Astronomy

Rensselaer Polytechnic Institute

Troy, NY 12180-3590, USA

(Submitted: 11 November 2017

Revised: 17 January 2018)

The statistical effects of different conditions in physical vapor deposition, such as sputter deposition, have on thin film morphology has long been the subject of interest. One notable effect is that of column development due to differential chamber pressure in the well-known empirical model called Thornton's Structure Zone Model. The model is qualitative in nature and theoretical understanding with quantitative predictions of the morphology is still lacking due to, in part, the absence of a quantitative description of the incident flux distribution on the growth front. In this work, we propose an incident Gaussian flux model developed from a series of binary hard-sphere collisions and simulate its effects using Monte Carlo methods and a solid-on-solid growth scheme. We also propose an approximate cosine-power distribution for faster Monte Carlo sampling. With this model, it is observed that higher chamber pressures widen the average deposition angle, and similarly increase the growth of column diameters (or lateral correlation length) and the column-to-column separation (film surface wavelength). We treat both the column diameter and the surface wavelength as power laws. It is seen that both the column diameter exponent and the wavelength exponent are very sensitive to changes in pressure for low pressures (0.13 Pa to 0.80 Pa); meanwhile both exponents saturate for higher pressures (0.80 Pa to 6.7 Pa) around a value of 0.6. These predictions will serve as guides to future experiments for quantitative description of film morphology under a wide range of vapor pressure.

---

[1] meesew@rpi.edu
[2] lut@rpi.edu

# I. INTRODUCTION

The implementation of thin films is commonplace in engineering to a wide range of technologies, including tribology, electronic devices, and optical coating[1–3]. The distinct physical properties of any particular film are often the product of the film's structure – that is, the film's morphology. Thus, the quantitative and statistical description of a film's morphology grown under specific deposition conditions is of theoretical interest and has been for some time[1,4–6]. In this work, we turn our attention to commonly employed sputter depositions.

Experimental effects of substrate temperature and sputtering pressure have long been documented and are summarized in Thornton's Structure Zone Model (SZM)[4,5]. In the region of the SZM where the temperature of the substrate $T$ is much less than the film's melting point $T_m$, $(T/T_m < 0.3)$[4], then we may neglect the kinetic effects produced by substrate temperature on the film's morphology, and we instead focus on the role the chamber pressure plays on the resulting structure within this low temperature region, known as "Zone 1". There is particular interest on how the change in pressure affects the film morphology seen in Thornton's seminal figure, where one sees the development of columnar structures with increasing argon pressure from 0.13 Pa to 4.00 Pa (Thornton worked in units of mTorr, where the equivalent pressure range is 1.0 mTorr to 30.0 mTorr)[4,5]. A qualitative sketch of this observed behavior at a fixed temperature is given in Fig. 1.

The use of Monte Carlo simulations to model the evolution of thin films has become common practice in the literature, as the growth of thin films is a problem far from equilibrium[1] and the conventional equilibrium statistical mechanics does not apply. However, when dealing with the chamber gas dynamics, which is far away from the growth front, it is common to approximate aspects of this non-equilibrium problem with known equilibrium phenomena. For example, many authors looking to simulate sputter depositions base their simulations on a particle's path from the target to the substrate during which the depositing particle may occasionally collide with a sputtering gas particle in the chamber. The distance the depositing particle travels before colliding with the gas is determined by Poisson statistics where the average distance a depositing particle travels is then given by the equilibrium value for the mean free path of the sputtering gas[7–11]. These simulations often employ algorithms to calculate energy distributions of collisions, and from there, the new direction a depositing species takes after each individual collision. Many of these simulations consider the effects of the emitted flux from the target have on the evolving film, while other authors are more interested in the effect of incident flux distributions have on surface morphology[1,3,12–15]. The term "particle flux" is to be interpreted as the normalized differential number of particles incident through a spherical differential solid angle[16]. In these works, the authors usually assume the deposition geometry is such that most particles travel without transport collisions from the target to the substrate; thus the incident flux distribution relative to the flat substrate is identical to that of the emitted flux distribution from the target. Table I shows the methods and some of the results relevant to this work of previously published simulations.

TABLE I: A brief collection of relevant works on flux distributions with some of their conclusions. The references to which each element belongs is given in the superscript. The first column names each particle flux distribution while the second column gives the probability distribution function (PDF) used in Monte Carlo sampling in terms of the polar angle $\theta$. The last row (*) shows our results for comparison.

| Flux distribution | PDF($\theta$) $\propto$ | Ejection or incident angle | Transport collisions | Brief summary of conclusions |
|---|---|---|---|---|
| Normal incident | $\delta(\theta)\sin\theta$ | Either[1,17]. | None[1,17] | Ballistic aggregation leads to KPZ growth[1,17]. |
| Cosine (isotropic) | $\cos\theta\sin\theta$ | • Incident[1,3,13,15,17]. <br> • Ejection[7,8]. | • None[1,3,13,15,17]. <br> • Poisson Process[7,8]. | Depending on the growth scheme, the wavelength exponent ranges from 0.5 to 1.0 [1,24,15]. |
| Cosine-power ($n$ is a fitting parameter) | $\cos^n\theta\sin\theta$ | • Ejection[7,8] | • Uniformly-Separated[19] <br> • Poisson Process[7,8] | The value of the exponent increases with the deposition's inverse Knudsen number[19]. |
| *Gaussian ($\sigma$ is a function of chamber pressure) | $\exp\left(-\dfrac{\theta^2}{2\sigma^2}\right)\sin\theta$ | Incident | Hard-sphere | Surface wavelength and column diameter exponents are sensitive to pressure from 0.13 Pa to 0.80 Pa, and then both saturate around 0.6. |

One common flux distribution in the literature is the "normal incident" distribution, where particles were modeled as traveling in a line-of-sight trajectory that is normal to a flat substrate. It was found that surfaces grown under a ballistic aggregation scheme with a normal incident flux distribution belong to the Kardar-Parisi-Zhang (KPZ) universality class[17]. In their simulations, they modeled the surface evolution without transport collisions. They also found that any deviation from the normally incident distribution leads to a breakdown of dynamic scaling[17]; thus the surface wavelength (measure of the column-to-column separation) and column diameters were shown to have unequal time dependence under normally incident particle flux.

The most readily used flux distribution for sputter simulations is the cosine flux, which is taken from the Knudsen Cosine Law used to describe isotropic particle flux[16]. As mentioned above, there has been two approaches to include particle flux in Monte Carlo simulations of thin film evolution: one that follows depositing particles from the target to the substrate, and the other that looks at the incident flux of particles already about to deposit at the substrate. An important result found from studying transport collisions is that the hard-sphere potential produces nearly identical thickness evolution as other intermolecular potential models when the particles are emitted from the target with a cosine flux distribution[7]. Another result found by Eisenmenger-Sittner while studying the transport collisions is that depositions under higher sputtering pressures led to the so-called "gas collimation effect" from the target, producing a higher proportion of oblique flux at the surface[8].

When studying the impact of the cosine flux distrbution on film morphology grown with a solid-on-solid growth scheme, Pelliccione *et al*. (2006) found that the surface wavelength exponent was relatively fixed at 0.5 for many different deposition conditions while the lateral correlation length exponent varied for those same conditions[1,3]. In that work, they also showed that dynamic scaling breaks down for high sticking coefficients, an effect attributed to wavelength selection due to mound formation[1,3]. Meanwhile, Lehnen and Lu (2010) found that surfaces grown under a cosine flux distribution with a ballistic aggregation growth scheme developed wavelength exponents of $0.67 \pm 0.04$ [17]. Alvarez *et al*. (2010) recorded a wavelength exponent of about 1.0 for films grown under a ballistic aggregation scheme, and their experimental results were verified by a Monte Carlo model[15]. The specific physical mechanism for why this value is substantially higher than for the others is unknown; however wavelength exponents of unity have been predicted in the past by considering random growth by shot noise limited by surface diffusion[18].

A fundamental assumption with the Knudsen Cosine Law is that the flux is isotropic. However, there exist systems where this assumption is no longer approximately true, and under those circumstances, cosine-power flux distributions have been used instead, where the actual power itself is fitted to experimental data[6,8,19]. In general, if a flux distribution has a power greater than unity, it is referred to as an *over-cosine* distribution, whereas a power less than unity is referred to as an *under-cosine* distribution. Eisenmenger-Sittner *et al*. (1995) observed that an over-cosine distribution is necessary to model particles ejected from the surface at higher pressures[8]. Furthermore, it was later found that the power of the over-cosine distribution at the target increases with the inverse Knudsen number, where the Knudsen number is the ratio of mean free path in the sputter chamber and the target diameter[19].

The authors emphasize, however, that the emitted particle flux at the target is not generally the same flux distribution at the growth front due to transport collisions between the target and the substrate. Thus, in this manuscript, we present a statistical model for the angular flux distribution that the depositing particles have before their first deposition along the evolving film due to transport collisions. We focus on the effect that binary hard-sphere collisions have on film morphology – that is, the collisions between the sputtering gas and the depositing species as the latter moves from the target to the substrate. From these collisions, we propose a statistical model for the flux distribution of the depositing particles at the surface. Based on this model, we employed a Monte Carlo algorithm known as the inverse cumulative distribution function (CDF) method[1], or simply the inversion method[20], to study the resulting simulated film morphology as a function of chamber pressure. Quantitative power law behaviors are predicted for the growth of column diameter and surface wavelength as a function of film thickness (growth time) under different chamber pressures.

## II. THEORETICAL METHODS

### A. Gaussian particle flux distribution

Consider a depositing specie constituting a classical ideal gas that travels from a target to the substrate, the latter of which is set as the origin of coordinates. As the film evolves, its height $h$ at any position $\boldsymbol{r} = (x, y, z)$ and time $t$ is denoted by $h = h(\boldsymbol{r}, t)$. Concurrently, we assume that, on average, the sputtering gas particles are separated by the gas' mean free path, $\lambda$. The mean free path of hard-sphere particles is known to be

$$\lambda = \frac{kT}{P\pi d^2 \sqrt{2}},\tag{1}$$

where $k$ is the Boltzmann constant, $T$ is the equilibrium temperature of the gas, $P$ is the equilibrium pressure of the gas, and $d$ is the particle diameter[21]. Thus, a depositing particle will be free to travel an average distance of $\lambda$ before colliding with the sputtering gas particle, assuming the sputtering gas density is much greater than the emitted particle density. A possible particle trajectory for these conditions is shown in an inset sketch in Fig. 2a.

Assuming the particles are emitted normally with respect to the substrate, they collide elastically with one another, and have equivalent masses, then there is an angular spread of $\pi/2$ between them after collision[22]. It is possible to obtain the average recoil angle $\Delta\theta$ for a particle assuming these collisions are isotropic, where $\theta$ is the polar angle measured relative to the substrate normal, which is taken to be the $z$-axis

$$\Delta\theta = \frac{2}{\pi}\int_0^{\pi/2}\left(\frac{\pi}{2} - \theta\right)\mathrm{d}\theta = \frac{\pi}{4},\tag{2}$$

as is shown in the inset sketch in Fig. 2b. The distance the average particle will travel towards the substrate will be

$$\langle l \rangle = \langle \lambda\cos\theta \rangle = \frac{2}{\pi}\int_0^{\pi/2}\lambda\cos\theta\,\mathrm{d}\theta = \frac{2\lambda}{\pi},\tag{3}$$

where $\theta$ is the (spherical) polar angle relative to the $z$-axis. The cosine function is needed to project the particle displacement onto the $z$-axis. The second to last equality follows from the assumed isotropy of the collisions. The average number of collisions $m$ a particle will undergo before depositing on a film is then

$$m = \text{floor}\left\{\frac{1}{\langle l \rangle}\left[L - h(\boldsymbol{r}, t)\right]\right\}, \tag{4}$$

where $L$ is the distance from the target to the substrate. Assuming $L \gg h(\boldsymbol{r}, t)$, $\forall \boldsymbol{r}$, where the symbol $\forall$ means "for all," then we may neglect the second term in $m$. We note that by neglecting the second term, however, necessarily implies this model describes film growth far outside of the atmospheric pressure regime modeled by Merkh and Lu[2], where antishadowing dominates surface evolution.

We note that this measure of the number of collisions is not exact, but rather better represents the average number of collisions in a Poisson process that any given particle undergoes before finally being deposited on the evolving film. A more general, albeit much more computationally expensive, approach is addressed in Appendix A. For the rest of this manuscript, we assume that each depositing particle will undergo the same number of collisions given by Equation (4).

We assume each collision has binary outcomes: one being the case were a particle is emitted with a recoil angle of $\Delta\theta$ with respect to the $z$-axis, while the other recoil angle is $-\Delta\theta$. We assume the probability of the former is $\Pr(\Delta\theta) = p$, while the latter case's probability is $\Pr(-\Delta\theta) = 1 - p$. By the assumed isotropy, the expectation value of the statistical first moment in $\Delta\theta$ vanishes. Meanwhile, the statistical second moment is given by

$$\langle \Delta\theta^2 \rangle = \sum_{k=1}^{m} \Pr(\Delta\theta_k)\,(\Delta\theta_k)^2 = \sum_{k=1}^{m}\left[p(\Delta\theta)^2 + (1-p)(-\Delta\theta)^2\right] = m\Delta\theta^2. \tag{5}$$

Therefore, the standard deviation throughout these average collisions is given simply by the root-mean-square of $\Delta\theta$

$$\sigma = \sqrt{\langle \Delta\theta^2 \rangle} = \Delta\theta\sqrt{m} = \frac{\pi}{4}\sqrt{m}\,. \tag{6}$$

Note that the value for $\sigma$ was derived assuming the sputtering gas particles and the depositing particles have the same masses. However, when this assumption does not hold, we may only consider two limits which are shown in Thornton and Marion[22]. The first is when the sputtering gas particles are much more massive than the depositing particles. In this case, the average recoil angle of the depositing particles $\Delta\theta$ can be calculated to be $\pi/2$ rather than $\pi/4$, as is shown by Equation 2. Computationally, the same effect can be achieved if $m \to 4m$ in Equation 6. Thus, in this limit, we may consider instead that the effective number of collisions increases by a factor of four. In the other limit, when the depositing particles are much more massive than the sputtering gas particles, then the sine of the maximum recoil angle of the depositing gas is approximately the ratio of the particle masses[22], $M_2/M_1$, where the smaller mass is obviously in the numerator. Thus, in this limit, the average recoil angle approaches zero as the maximum recoil angle does. So we take the half of maximum to be the average recoil angle (as can be seen in Thornton and Marion, in this limit the recoil angles are coupled transcendental equations in terms of the masses[22], and therefore the average integral in Equation 2 cannot be evaluated analytically). With this newly defined recoil angle, the rescaling of $\sigma$ can instead be achieved by rescaling $m$ by $(2M_2)^2/(\pi M_1)^2$ in Equation 6. Hence, it suffices to continue with the model where particle masses are equivalent, and then rescale the variables of interest that depend on the number of collisions by an appropriate numerical factor.

After $\sigma$ has been determined, it is then possible to represent this angular distribution then as a Gaussian centered at zero with width $\sigma$. We conclude that the differential number of particles depositing through the differential surface area at $\theta$ per unit time is proportional to this Gaussian distribution; hence we can write the normalized particle flux $d\Phi$ through the differential solid angle $d\Omega$ between $(\theta, \phi)$ and $(\theta + d\theta, \phi + d\phi)$ as

$$d\Phi = \frac{1}{N} \exp\left(-\frac{\theta^2}{2\sigma^2}\right) d\Omega = \frac{1}{N} \exp\left(-\frac{\theta^2}{2\sigma^2}\right) \sin\theta \ d\phi d\theta \,, \qquad (7)$$

where $N$ is the normalization constant, $\theta$ is the polar angle, and $\phi$ is the azimuthal angle. We further assume that this particle flux is azimuthally symmetric. Then we can define the probability distribution $g(\theta, \sigma)$ by

$$g(\theta, \sigma) \equiv \frac{1}{Z} \sin\theta \exp\left(-\frac{\theta^2}{2\sigma^2}\right), \qquad (8)$$

where the normalization functional $\mathbb{Z}(\sigma)$ is given by

$$\mathbb{Z}(\sigma) = \int_0^{\frac{\pi}{2}} \exp\left(-\frac{\theta^2}{2\sigma^2}\right) \sin\theta \ \mathrm{d}\theta. \qquad (9)$$

It is interesting from a computational point-of-view to note that $\mathbb{Z}$ has an analytic form, and it is derived in Appendix B. In the meantime, we focus our attention on approximating $g(\theta, \sigma)$ with a $\cos^n \theta$ flux distribution, and more specifically, obtaining an explicit expression for $n$ in terms of $\mathbb{Z}(\sigma)$.

### B. Approximation with a $\cos^n \theta$ distribution

It may be necessary, with respect to computational efficiency, to use a different, simpler flux distribution to sample the polar angle of depositing particles for simulations. One such distribution is the cosine-power distribution, $\cos^n \theta$. We note that this is not the same cosine-power distribution often used to model the emitted polar angle from a target, as described in [6,19,23]. The approximate cosine-power distribution of particular interest here is instead used to describe angular spread before deposition at the growth front after a finite series of discrete, hard-sphere collision events. Using the inversion method for Monte Carlo random sampling, one can more easily sample polar angles from a uniform random number $x \in [0, 1]$ by

$$x = (n+1) \int_0^\theta \cos^n \theta' \sin\theta' \ \mathrm{d}\theta' \Rightarrow \theta(x) = \cos^{-1}\left(\sqrt[n+1]{1-x}\right), \qquad (10)$$

where the prefactor of $(n+1)$ constitutes the normalization constant for the cosine-power distribution. It is clear that explicit inversion in terms of elementary functions is not possible with the Gaussian flux distribution given by Equation (8) and its cumulative distribution function (CDF) given in the Appendix AII. However, an approximate value for the exponent can be obtained, given any chamber pressure, so that an approximate distribution can be used in lieu of Equation (8) – as the latter necessarily requires a more robust, hence more computationally expensive, Monte Carlo sampling algorithm.

Before we address the $\cos^n \theta$ distribution, we need to establish a couple of properties associated with $\mathbb{Z}$. In this section, we choose to treat $\sigma$ as a continuous variable on the interval $(0, \infty)$ for the purposes of analyzing the asymptotics of $\mathbb{Z}$. We demonstrate below that $\mathbb{Z}(\sigma)$ is monotonically increasing on the interval $(0, 1)$.

To prove that $\mathbb{Z}(\sigma)$ is monotonically increasing, it suffices to show its first derivative is a nonnegative function.

$$\begin{aligned}
\frac{d\mathbb{Z}}{d\sigma} &= \int_0^{\pi/2} \left[ \frac{\partial}{\partial \sigma} \exp\left( -\frac{\theta^2}{2\sigma^2} \right) \right] \sin\theta \ d\theta \\
&= \frac{1}{\sigma^3} \int_0^{\pi/2} \theta^2 \exp\left( -\frac{\theta^2}{2\sigma^2} \right) \sin\theta \ d\theta \\
&= \left| \frac{1}{\sigma^3} \right| \int_0^{\pi/2} \left| \theta^2 \exp\left( -\frac{\theta^2}{2\sigma^2} \right) \sin\theta \right| d\theta \geq 0 \,.
\end{aligned}$$

The inequality follows from the definite integral of a nonnegative function and that $\sigma \in (0, \infty)$. To show that $\mathbb{Z}(\sigma) : (0, \infty) \to (0, 1)$ is a one-to-one and onto function, it suffices to establish the two extreme limits of $\mathbb{Z}(\sigma)$ and then rely on its monotonicity.

$$\lim_{\sigma \to 0^+} \mathbb{Z}(\sigma) = \int_0^{\pi/2} \left[ \lim_{\sigma \to 0^+} \exp\left( -\frac{\theta^2}{2\sigma^2} \right) \right] \sin\theta \ d\theta = \int_0^{\pi/2} (0) \sin\theta \ d\theta = 0$$

$$\lim_{\sigma \to \infty} \mathbb{Z}(\sigma) = \int_0^{\pi/2} \left[ \lim_{\sigma \to \infty} \exp\left( -\frac{\theta^2}{2\sigma^2} \right) \right] \sin\theta \ d\theta = \int_0^{\pi/2} (1) \sin\theta \ d\theta = 1 \,.$$

With these properties of $\mathbb{Z}(\sigma)$ established, we proceed with the analysis with the $\cos^n \theta$ flux distribution.

We define a "fitting" value of $n$ to the Gaussian distribution as the exponent of the $\cos^n \theta$ which reduces the average difference between the $\cos^n \theta$ distribution and the Gaussian distribution to a finite value less than $2\epsilon/\pi$ for a given $\epsilon > 0$ and some Gaussian width $\sigma$. Written explicitly,

$$\frac{2}{\pi} \int_0^{\pi/2} \left| \cos^n \theta - \exp\left( -\frac{\theta^2}{2\sigma^2} \right) \right| \sin\theta \ d\theta < \frac{2\epsilon}{\pi} \,. \tag{11}$$

It is noted that this process is equivalent to reducing the area between the two flux distributions to an arbitrarily small value. Since the sine is nonnegative over this nonnegative partition, it is possible to rewrite this inequality as

$$\left| \int_0^{\pi/2} \cos^n \theta \sin \theta \ d\theta - \int_0^{\pi/2} \exp\left(-\frac{\theta^2}{2\sigma^2}\right) \sin \theta \ d\theta \right| = \left| \frac{1}{n+1} - \mathbb{Z}(\sigma) \right| < \epsilon \,. \tag{12}$$

Therefore, for every $\mathbb{Z}(\sigma)/2 > \epsilon > 0$, we can find an interval of $n$ given by

$$\frac{1}{\mathbb{Z}(\sigma) - \epsilon} - 1 > n(\sigma) > \frac{1}{\mathbb{Z}(\sigma) + \epsilon} - 1 \,. \tag{13}$$

Finally, in the limiting case in which $\epsilon \to 0$, we obtain the "best-fitting" value of $n$ – that is the exponent for which the average difference between the two distributions vanishes:

$$n(\sigma) = \frac{1}{\mathbb{Z}(\sigma)} - 1 \,. \tag{14}$$

By combining Equations (1), (4), and (6), while ignoring the floor function in Equation (4), it is clear that $\sigma \propto P^{1/2}$, a monotonically increasing relationship when one holds all other parameters in Equation (1) constant, and therefore $\mathbb{Z}(P)$ increases with increasing pressure, implying $n(\mathbb{Z})$ decreases with increasing pressure. One can interpret this result as a statement that increasing the number of collisions in the sputter chamber will produce an increasingly uniform particle flux at the surface. In this context, "uniform" does not mean "isotropic" as would be the case with the Knudsen Cosine Law. Instead, uniformity in $\theta$ implies that the differential particle flux through a differential solid angle at $\theta$ is constant. Meanwhile, we note there is a one-to-one correspondence between continuous pressure $P$ and the best-fitting value of $n$. Since both $\sigma(P)$ and $\mathbb{Z}(\sigma)$ are monotonically increasing, it suffices to show $n(\mathbb{Z})$ is monotonic and lies on the interval $(0, \infty)$. As before, we establish the monotonicity first.

$$\frac{dn}{dZ} = -\frac{1}{Z^2} < 0, \forall Z \in (0,1).$$

This implies that $n$ is monotonically decreasing. It is then clear that

$$\lim_{Z \to 0} n(Z) = \infty, \qquad \lim_{Z \to 1} n(Z) = 0\,.$$

Finally, since $Z(P)$ is a one-to-one and onto mapping from $P \in (0, \infty)$ to $Z \in (0,1)$, $n(P)$ is a one-to-one and onto mapping from $P \in (0, \infty)$ to $n \in (0, \infty)$, for continuous $P$. Outside of the continuum limit, this one-to-one correspondence implies there is a unique best-fitting value of $n$ for any given number of collisions $m$ between the target and film (outside the continuum limit, $m$ obtains discontinuities due to the floor function).

## C. Monte Carlo methods

The numerical calculations performed consisted of $(2 + 1)$-dimensional Monte Carlo simulations on a $N \times N \times L$ simple cubic lattice with periodic boundaries, where $N = 512$ and $L = 1024$. A brief summary of typical simulations used is as follows.

After the lattice is created and the time step $t$ and particle number $p$ are each set to one, then particles are created and move inside the lattice. After a total of $R = 62500$ particles are deposited, then one is added to $t$ until $t = t_{max}$, where $t_{max}$ is the total number of simulations steps in each simulation. For all simulations, $t_{max} = 1000$. Meanwhile, each particle of a total of $Rt_{max}$ undergoes the following deposition algorithm.

First, the particle's initial position is randomly selected in the $xy$-plane, while the $z$-coordinate is simply selected as one lattice unit above the maximum height of the evolving surface. We note that this selection is statistically equivalent to choosing any other greater initial $z$-coordinate, as is shown in Appendix C. Then, the azimuthal angle of the particle's trajectory $\phi$ is sampled from a uniform distribution such that $\phi \in [0, 2\pi)$ while the polar angle $\theta$ must be sampled via the distribution given by Equation (8) using the inversion method. From there, the particle travels to the surface, where a particle may stick to a surface given a sticking probability given by the zeroth order sticking coefficient $s_0 \in [0, 1]$. To stick to the surface, a depositing particle must come close enough to it so that at least one of the already-occupied surface lattice points is within the set of nearest or next-nearest points of the depositing particle (in a cubic lattice, there are 26 of these locations around any given point in the lattice). If the particle does not stick, then it is reemitted via a specular reflection law. Normally, sticking coefficients of all orders could exist, however, for simplicity in these simulations, $s_0 = 0.65$ and $s_1 = 1$. The value of $s_0$ was chosen to be consistent with previous simulations modeling silicon film evolution[1]. Once the particle finally "sticks" to the surface at $(x, y, z)$, a solid-on-solid model of film growth is employed, during which the particle height is automatically set to be $h + 1$, where $h = h(x, y)$ is the height of the simulated film at position $(x, y)$.

Next, a Boltzmann diffusion algorithm is run over $D/R$ individual events, wherein $D$ is interpreted as the "strength" of diffusion and $R$ is aforementioned number of particles per simulation step. After the particle is deposited, $D/R$ nearby particles are randomly selected, and each is assigned an energy of $E = E_a + CE_b$, where $E_a$ is the activation energy for the particles, $E_b$ is the bond energy between particles, and $C$ is the number of adjacent particles. Each of the selected particles to undergo diffusion are then assigned a random number $\xi \in [0, 1]$. If $\xi \leq B \equiv \exp[-E/(kT_s)]$, where $k$ is the Boltzmann constant and $T_s$ is the temperature of the substrate, then the selected particle is moved to a randomly selected, unoccupied location in the set of 26 nearest or next-nearest particle locations that is consistent with the solid-on-solid model. After $D/R$ diffusion events are undergone, then the simulation proceeds with the next particle. For these simulations, $D/R = 100$, $E_a = 0.08$ eV, $E_b = 0.05$ eV, and $T_s = 298$ K, so that this deposition algorithm's parameters are consistent with those that produce silicon-like simulated films in the literature[3]. A schematic of the simulation algorithm is shown in Appendix D.

The specific independent variable tested in these simulations was the chamber pressure $P$. The other three variables in Equation (1) were held constant throughout these simulations and had values of $T = 300$ K, and $d = 2.12$ Å. The latter value was chosen as the covalent diameter of argon – a value between the atomic diameter and the van der Waals diameter for argon. These values approximately corresponded to one collision at a pressure of 0.13 Pa (1.0 mTorr), when the target-substrate distance was taken to be a fixed value of $L = 10$ cm.

The dependent variables studied were the exponents associated with lateral scaling behavior, namely the inverse dynamic exponent $1/z$ and the wavelength exponent $p$. The former describes the scaling behavior of the film's lateral correlation length $\xi(t) \sim t^{1/z}$ while the latter describes that of the film wavelength $\lambda(t) \sim t^p$. Here we take the lateral correlation length to be the distance at which the film's normalized autocorrelation function is $e^{-1}$, and the film wavelength is the peak wavelength given in the power spectral density function of the film[1]. The computational methods used to obtain these values are sketched in Fig. 4a and 4c. Thus the lateral correlation length is taken to be a measure of an average column's diameter, while the wavelength is taken to be a measure of the average distance between adjacent columns. In earlier work, it has been noted that for low pressure solid-on-solid models grown from a cosine flux distribution, $p = 1/2$ in the absence of diffusion[1,3,24], while $1/z = 1/2$ [25]. Notably, in the same work it is shown that if a film obeys dynamic scaling relationships, then $p = 1/z$. For these films, there is no characteristic length scale that develops over time; instead there is a relative length scale $\lambda(t)/\xi(t) \equiv c(t^p/t^{1/z}) = c$, where $c$ is a constant[1,3]. In mounded surfaces, where dynamic scaling breaks down, then $p$ is observed to be universal throughout many different simulated deposition conditions whereas $1/z$ varies significantly with changes in deposition conditions[1,3]. One of purposes of this work is to better understand how the number of collisions induced by higher pressures would lead to deviations in the wavelength exponent universality in low pressure depositions.

## III. RESULTS AND DISCUSSIONS

### A. Pressure-dependent flux distributions

Figure 2a shows a plot of the number of collisions a gas will undergo as a function of pressure, according to Equation (4) under the assumption that $L \gg h(r, t)$. As can be deduced from the plot, the minimum pressure whose Gaussian flux distribution can be examined is the inverse of the slope, namely 0.13 Pa (or 1.0 mTorr), since this pressure corresponds to $m = 1$. At the same time, with the chamber parameters chosen as they are, Equation (4) is an approximate identity mapping between integer chamber pressures measured in millitorr to the number of collisions that occur between the target and the substrate for comparison to Thornton's SZM.

The Gaussian flux distribution is sensitive to the chamber pressure, or equivalently, the number of collisions a particle undergoes before deposition. Figure 3 shows a few properties of this flux distribution as a function of the chamber pressure, the most obvious of which is that the shape is very sensitive to changes in pressure for lower pressures, but are not very sensitive to changes in pressure for higher pressures. The Fig. 3a shows how the flux distribution curve changes for different pressures. Note how the flux distribution very quickly approaches a normalized sine curve – a polar angle distribution that arises from a uniform probability distribution. Fig. 3b and Fig. 3c show the mean polar angle as a function of chamber pressure and the root-mean-square (rms) angle as a function of chamber pressure, respectively. Both plots show, too, that these values quickly approach that of the uniform distribution, namely 57.3° and 61.2°, respectively. Figure 3d shows the width of the Gaussian flux distribution as a function of chamber pressure, where the width $w$ of a function whose argument is $x$ is defined by the usual relation $w^2 \equiv \langle (x - \langle x \rangle)^2 \rangle = \langle x^2 \rangle - \langle x \rangle^2$. This plot similarly approaches its sinusoidal limit for high pressures (21.56°), however, the width lacks monotonicity when both the mean angle and the rms angle are monotonically increasing with pressure, where the crossover chamber pressure can be seen to be near 0.27 Pa.

Using Equation (14), the best-fitting $n$-values from the $\cos^n \theta$ flux distribution were calculated while varying the chamber pressure. These exponents are plotted in the rightward plot of Fig. 2b. As can be seen in the plot, there are three domains where $\log n$ is proportional to $\log P$. The linear fits over these domains are superimposed on the plots and their respective curves on linear axes are shown in the figure's legend. We note that the $R^2$ values calculated on the log-log axes substantiate the least-squares values shown, and show there is very little error in the approximate mapping between chamber pressure and $n$-value for quicker implementation in Monte Carlo sampling.

## B. Monte Carlo simulation results

The film wavelength $\lambda$ and lateral correlation length $\xi$ were measured as functions of time and pressure from the simulated films. Figures 4a and 4c show the films' wavelengths and lateral correlation lengths as functions of simulation time steps, for all the varying pressures. In these plots, it is clear that lower pressures will tend to produce lower wavelengths and lateral correlations lengths in time. The higher pressures then all tend to saturate around the average slope shown by the black dotted line whose numerical values are shown in the plots' legends. The respective growth exponents, $p$ such that $\lambda(t) \sim t^p$ and $1/z$ such that $\xi(t) \sim t^{1/z}$, were extracted from each of these curves on log-log axes, where the exponent's value is given by the slope of the line of best fit. The values for these growth exponents were plotted as functions of pressure in Fig. 4b and 4d, where the error bars were calculated from the covariance matrix generated by the curve-fitting algorithm used. As can be seen from the figure, there is a fairly linear or constant relationship between the value of these exponents and the chamber pressure on semilog axes, and the least-squares curves are given in the legend of each plot. For lower pressures ($P \leq 0.80$ Pa), the exponents appear to grow linearly with pressure, and this linear fit is reasonably justified by the $R^2$ values found for $p(P)$ and $1/z(P)$ – those values being about 0.93 and 0.88, respectively. For higher pressures ($P > 0.80$ Pa), the growth exponents seem to saturate around the values of $p = 0.6112 \pm 0.0237$ and $1/z = 0.6093 \pm 0.0154$.

Figure 5 shows a few plots of the lateral correlation length $\xi$ as a function of pressure, at a specific height. The height for each plot is given as the mean film height, and it is reported in each plot as the title. The error in the mean height comes from the standard deviation of the heights of the films grown at the different pressures. It is noted that in each plot, again lower pressures correspond to smaller film wavelengths while the higher pressures correspond to an apparent wavelength saturation, regardless of the height. Meanwhile, the saturation level increases with pressure. Furthermore, there are two identifiable domains: low pressures ($P \leq 0.80$ Pa) and high pressures ($P > 0.80$ Pa). Over these two domains, the lateral correlation length is linear in $\log_{10} P$, particularly at lower film heights, and the respective curves are given in the legend of each plot. Another pattern that is shared by all the sampled heights is that, the linearity breaks down as the films get thicker. This is verified by the gradual decrease of the calculated $R^2$ values as the mean height increases. A similar set of figures was attempted for the wavelength, but due to the larger variation between the $\lambda(t)$ plot and the $\xi(t)$ plot in Fig. 4, there were no clear trends for $\lambda(P)$ at a fixed height other than the one already discussed.

A visual representation of how the film's lateral correlation length is a function of pressure can be seen in Fig. 6. There are a series of lateral cross-sections created by the simulation. The black regions represent empty space, while the lighter pixels represent occupied lattice boxes. The darker gray regions represent occupied lattice boxes, but behind the plane through which we image the cross section. The closest plane is completely shaded in to help further contrast the inter-column separation (or lack thereof) in the images. It can be seen in the figure that for lower pressures, the film has very little lateral correlation, hence the rod diameter is very small. This effect leads to the relative uniformity in the leftward images in contrast to the rightward images. As the pressure gets higher, columns quickly begin to appear along the surface and widen with increasing pressure, but then after about 0.93 Pa, the column diameters seem to be about the same size for the two heights shown, in accordance with the top two plots in Fig. 6. We note that this effect is only exaggerated as the films continue to grow in time, as is shown in Fig. 7, where one can see that surface features are much more clearly defined for the cross section of the film at 4.00 Pa at later time steps.

## IV. DISCUSSIONS AND CONCLUSION

In this work, we examine a new computational model to study how sputter depositions at different chamber pressures will affect the surface morphology of films belonging to Thornton's Zone 1 in his Structure Zone Model[5]. In particular, this new model is proposed as a predictive means to translate changes in chamber pressure into changes in how the wavelength and lateral correlation length evolve in time.

The Gaussian flux distribution in Equation (8) is built from a series of hard-sphere collisions a depositing specie undergoes as it travels from the target to the substrate. This model assumes that the individual collisions are isotropic and occur in the exact positions in the chamber as would be expected for the sputtering gas comprised of a lattice of characteristic length given by the average projection of the gas' mean free path on the $z$-axis (see Equation (3)). Such a configuration is statistically unlikely, however, Equation (8) can be thought of as the most probable flux distribution arising from a more general flux distribution, where the number of collisions themselves, and by extension the width of the Gaussian, is randomly selected from a Poisson distribution whose average is given by $L/\langle l \rangle$, where $L$ is the distance between the target and the substrate and $\langle l \rangle$ is the average vertical distance a particle moves before colliding with the sputtering gas.

The calculation of the individual angular moments of the Gaussian flux distribution as in Fig. 3b, 3c, and 3d show that the shape of the distribution is very sensitive to changes in pressure at low pressures, but then the moments rather quickly approach those of a uniform flux distribution as the pressure increases. This effect of pressure was consistent throughout all of the measured data, wherein the measured quantities were sensitive to pressure for low pressures, but then saturated at higher pressures.

It has been shown that an approximate cosine-power distribution exists as a substitute for the Gaussian flux distribution to speed the Monte Carlo inverse CDF sampling in simulations. An explicit value for the power can be calculated given any nonzero number of collisions, as is shown by Equation (14). It was later shown that the exponent of the cosine-power distribution is inversely proportional to the chamber pressure, and least-squares regression curves are given to quickly compute the value exponent if a simulation were to be written using this approximate flux distribution. This further indicates that increasing the chamber pressure increases the uniformity in the polar angles any particular depositing particle may have.

This approximation with the cosine-power distribution appears to hold reasonably well when it is checked against known data tabulated in[1]. For example, to model a cosine flux distribution, the exponent would be one, and by using the least-squares regression curve over the first domain in Fig. 2b, the corresponding chamber pressure would be 0.110 Pa. By further extrapolating from the least-squares curves, the corresponding wavelength exponent and lateral correlation length exponent are $p(0.110 \text{ Pa}) = 0.436$ and $1/z(0.110 \text{ Pa}) = 0.475$, respectively. Comparing these values to those in Table 8.1 in[1] or Table I in[3] for an initial sticking coefficient of $s_0 = 0.625$ with a diffusion strength of $D/R = 100$, the wavelength exponent $p$ is within error $(0.48 \pm 0.04)^{1,3}$ to two significant figures, but the value of $1/z$ is higher than expected $(0.40 \pm 0.03)^{1,3}$. However, as noted in both works, the values of $p$ typically tend to be universal while the value of $1/z$ is sensitive to the specific deposition conditions used for solid-on-solid growth schemes. Perhaps the discrepancy in $1/z$ is a result in the error between the Gaussian flux distribution employed in this work and the cosine distribution used in[1,3].

An interesting effect that was observed in these simulations is that $p$ and $1/z$ saturate at high pressures around the value of 0.6. Therefore, high pressures seem to induce an "anomalous scaling" effect between the lateral correlation length and the wavelength, despite there being clear wavelength selection due to the development of mounds. The authors note that the saturated value for lateral correlation length exponent is significantly higher than that found in solid-on-solid models simulated with similar deposition parameters and a cosine flux distribution[1,3]. This effect could be attributed to the Gaussian flux distribution providing more particles incident at angles higher than 45°, which would lead to faster lateral column growth, especially since for high pressures, the mean angle of incidence is greater than 45°. This idea and its effect is consistent with the results in[8], where particle depositions at higher pressures were incident with larger angles than those at lower pressures. Meanwhile, the lateral growth was observed to be limited by our simulations' programmed lack of overhangs, thereby preventing any fan-like shapes from dominating the column growth. Perhaps the competition between the increased lateral flux and the solid-on-solid growth scheme can explain why some authors observe or simulate much higher values for $1/z \approx 1.0$ for surfaces that do grow with allowable overhangs, as can be seen in[15]. The relationship between $1/z$ for films grown under a solid-on-solid scheme and those grown under a ballistic aggregation scheme is left as a topic of future work.

Table II compares a few experimental results reported in the literature to the Gaussian flux model's predictions, given a particular experimental setup. Since the Gaussian flux distribution has a fairly sharp transition between low pressures, but then saturates for high pressures, we would expect that the columnar structure observed for high pressures should saturate as well, and particularly if the Gaussian flux model was the dominant mechanism involved in sputtering. If this were true, then the wavelength exponent and column diameter exponent would each be 0.6, as predicted. However, Messier *et al.*[25] found that the wavelength exponent was usually $p = 0.5$ when the column diameters are power laws of film thickness for sputtering pressures at the higher end that we tested in this manuscript. They did note though that the value depended on the ion bombardment and substrate temperature (the method for taking the "average column diameter" at the time was by finding the characteristic wavelength using a Fourier transform, thus the exponent they report is $p$ rather than $1/z$). Meanwhile, Alvarez *et al.*[15] found that, at much lower sputtering pressures than Messier, the wavelength exponent and column diameter exponents were about unity. As aforementioned, it may be possible to explain Alvarez' high exponent by accounting for shot noise during the deposition process, as the shot noise scales the Fourier transform linearly in film thickness[18].

TABLE II: A comparison of experimental work with the predictions from our model of the deposition angle distribution given the deposition conditions.

| Relevant Work | Chamber Pressure | Target Source | Chamber Gas(es) | Observed Column Structure | Gaussian Flux Model Predictions |
|---|---|---|---|---|---|
| Messier *et al.*[25] (Radio Frequency Sputtering) | $2.6 - 5.2$ Pa | Si | Ar | $p = 0.5$* | $p = 0.61 \pm 0.02$ $1/z = 0.61 \pm 0.01$ |
| Pelliccione *et al.*[3] (DC Magnetron Sputtering) | 0.27 Pa | Si | Ar | $p = 0.51 \pm 0.03$ $1/z = 0.38 \pm 0.03$ | $p = 0.54 \pm 0.05$ $1/z = 0.56 \pm 0.01$ |
| Alvarez *et al.*[15] (Reactive Magnetron Sputtering)** | 0.5 Pa | Ti | Ar:O 10:1.3 | $p \approx 1.0$ $1/z \approx 1.0$ | $p = 0.60$ $1/z = 0.60$ |

\*   This value was found to depend on the substrate temperature and ion bombardment.
\*\*  In this work there is no explicit difference made between the column diameter exponent and
     the wavelength exponent, thus we take them to be equivalent.

An explanation for why the Messier exponent is somewhat lower than the predicted value for the same pressure and the observed value at lower pressures[15] is most likely multi-faceted. The first explanation is that the transport distance between the target and the substrate is much lower than the 10 cm distance we accounted for in this manuscript. However, a description of the transport distance is not given in the Ref. 25. Another explanation is that the exponents Messier found might be altered by other physical processes as well, and the lower value can be attributed to an effect the Gaussian flux model cannot completely account for on its own. The most likely source of the discrepancy is the higher temperature regime Messier was interested in describing within Thornton's SZM[25], and a theoretical description of the transition from a higher exponent to a lower exponent as a function of increasing substrate temperature could be a subject of future work.

Meanwhile, at lower pressures, the wavelength exponent predicted from the Gaussian flux model is in good agreement with that found experimentally by Pelliccione *et al.*[3] We note that this exponent is also in good agreement with the value calculated from the simulations in the same work[3], as described above. The experimentally observed column diameter exponent is somewhat lower than the one that is predicted by the Gaussian flux model. As mentioned in their work, while the wavelength exponent was conjectured to be universal, the value of column diameter exponent may be affected by other conditions during deposition, such as the sticking coefficient which we have not explored in detail.

In summary, it is observed that for films in Thornton's Zone 1, column widths depend on sputtering gas pressure. We proposed a statistical model for the incident angular flux distribution as a function of transport collisions between the depositing particles and the ambient sputtering gas, and we studied the resulting affects the flux distribution on film morphology using Monte Carlo simulations. We also proposed an approximate cosine-power flux distribution for easier implementation in future simulations. It was observed that our Gaussian flux model produced surfaces whose column widths and surface wavelengths were sensitive to pressures at low pressures, and then would saturate around the value $p \approx 1/z \approx 0.6$. The behavior of the exponents' numerical values mimics that of the polar angular moments of the proposed Gaussian flux distribution, and given that as this model's extrapolation returns exponent values similar to those known for simulations based on a cosine flux, we surmise that this model of sputter chamber pressure could be a liaison between the well-documented cosine flux distribution and a uniform flux distribution.

## ACKNOWLEDGEMENTS

The authors would like to thank Dr. G.-C. Wang for her help and advice throughout this project. We would also like to thank Mr. T. Merkh for valuable discussions. This work was supported in part through Rensselaer Polytechnic Institute's Undergraduate Research Program.

### APPENDIX A: A MORE GENERAL POISSON DISTRIBUTION

The motivation for deriving a more general Gaussian flux distribution comes from the fact that the sputtering gas in the chamber is not a lattice. We expect that the most probable number of collisions and the average number of collisions a particle will go through is given by Equation (4), however this approach necessarily creates discontinuities in the flux distribution employed in this manuscript. Specifically, this approach necessitates that the width of the Gaussian factor in Equation (8) be a function of nonnegative integers, thereby limiting that model's ability to probe a pressure continuum.

To remedy this problem, we eliminate the floor function in Equation (4) completely. By doing so, the number of collisions is no longer guaranteed to be a nonnegative integer; however, we employ the well-known theorem that the set of rational numbers are dense in the set of reals. Therefore, any real number $m$ given by

$$m \equiv \frac{1}{\langle l \rangle}[L - h(\boldsymbol{r}, t)] \approx \frac{L}{\langle l \rangle}, \qquad \forall \boldsymbol{r}, t : L \gg h(\boldsymbol{r}, t), \tag{A1}$$

can be approximated arbitrarily closely by two rational numbers, $p$ and $q$ such that $p \leq m \leq q$ where equality holds if and only if $m$ is rational. Suppose the lower bound $p$ is chosen as a sufficient approximation of $m$. Since $p \geq 0$ is rational, there exists two integers $a \geq 0$ and $b > 0$ such that $p = a/b$. Therefore, $m \approx a/b$. We interpret this ratio in the context of hard-sphere collisions by saying that we expect $a$ total particle collisions for every $b$ particles that are deposited on the surface. For example, suppose $m = 1.1$. Then we may interpret this result as there are 11 collisions for every 10 particles deposited.

With the assumption that the number of collisions any depositing particle goes through is independent of those undergone by the particles that were deposited before it, then the set of collisions a particle goes through is a Poisson process. Therefore, given a value $m$ calculated from Equation (A1), the probability that the depositing particle undergoes and integer $k \geq 0$ collisions is given by

$$\Pr(k; m) = \frac{m^k e^{-m}}{k!}. \tag{A2}$$

And therefore, the expected angular width from that $k$ collisions would be $\sigma_k = \left(\frac{\pi}{4}\right)\sqrt{k}$, by Equation (6), making the resulting (non-normalized) Gaussian flux distribution have the shape

$$g_k(\theta, \sigma_k) = \sin\theta \exp\left(-\frac{\theta^2}{2\sigma_k^2}\right). \tag{A3}$$

Thus, by summing over all $g_k$, $k \geq 0$, where each term is multiplied by its Poisson probability in Equation (A2), we recover the generalized Gaussian flux distribution $\tilde{g}$:

$$\tilde{g}(\theta, m) = \frac{1}{\tilde{\mathbb{Z}}} \sin\theta \sum_{k=0}^{\infty} \frac{m^k}{k!} \exp\left[-\frac{1}{k}\left(\frac{8}{\pi^2}\right)\theta^2\right]. \tag{A4}$$

where the generalized normalization functional $\tilde{\mathbb{Z}}$ can be written in terms of the single Gaussian normalization constant $\mathbb{Z}$ as

$$\tilde{Z}(m) \equiv \sum_{k=0}^{\infty} \frac{m^k}{k!} \int_0^{\frac{\pi}{2}} \exp\left[-\frac{1}{k}\left(\frac{8}{\pi^2}\right)\theta^2\right] \sin\theta \ \mathrm{d}\theta = \sum_{k=0}^{\infty} \frac{m^k}{k!} \ Z\left(\frac{\pi\sqrt{k}}{4}\right). \tag{A5}$$

Consequences of this generalized flux distribution could be a line of future study. In particular, this generalized flux distribution will allow a continuum of chamber pressures to be modeled. Similarly to how the Gaussian flux distribution in Equation (8) may serve as a liaison between the cosine flux distribution and a uniform flux distribution through varying chamber pressure, this generalized flux distribution may serve as a theoretical liaison between a normal flux distribution when $m \to 0$ and a uniform flux distribution when $m \to \infty$.

## APPENDIX B: AN ANALYTICAL EXPRESSION FOR Z

When implementing the Gaussian flux distribution in simulations, it might be more computationally efficient to write a function to determine the value of Z, given some Gaussian width $\sigma \geq 0$, rather than obtaining the value by numerical integration. Thus, it is beneficial to find an explicit representation of Z as a function of $\sigma$.

We begin by substituting the Euler expression for the sine function into Equation (8). By defining the imaginary unit as $i \equiv \sqrt{-1}$, we have

$$Z(\sigma) = \frac{1}{2i}\left[\int_0^{\pi/2} \exp\left(-\frac{\theta^2}{2\sigma^2} + i\theta\right)\mathrm{d}\theta - \int_0^{\pi/2} \exp\left(-\frac{\theta^2}{2\sigma^2} - i\theta\right)\mathrm{d}\theta\right]. \tag{B1}$$

We elect to define the integral $K_{\pm}$ by

$$K_{\pm} \equiv \int_0^{\frac{\pi}{2}} \exp\left(-\frac{\theta^2}{2\sigma^2} \pm i\theta\right)\mathrm{d}\theta \ . \tag{B2}$$

By completing the square, we rewrite $K_{\pm}$ as

$$K_{\pm} = \exp\left(-\frac{1}{2}\sigma^2\right) \int_0^{\pi/2} \exp\left[-\frac{1}{2\sigma^2}(\theta \mp i\sigma^2)^2\right] \mathrm{d}\theta \ . \tag{B3}$$

It becomes possible to integrate this function with the change of variable

$$s_{\mp} \equiv \frac{\theta \mp i\sigma^2}{\sigma\sqrt{2}}. \tag{B4}$$

Thus we finds

$$K_{\pm} = \sigma\sqrt{2}\, e^{-\frac{1}{2}\sigma^2} \int_{s_{\mp}(0)}^{s_{\mp}(\pi/2)} e^{-s_{\mp}^2}\, ds_{\mp} = \sigma\sqrt{\frac{\pi}{2}}\, e^{-\frac{1}{2}\sigma^2} \left[ \mathrm{erf}\left( \frac{\frac{\pi}{2} \mp i\sigma^2}{\sigma\sqrt{2}} \right) - \mathrm{erf}\left( \frac{\mp i\sigma}{\sqrt{2}} \right) \right]. \tag{B5}$$

Here we have made use of the definition of the error function, given by

$$\mathrm{erf}(t) = \frac{2}{\sqrt{\pi}} \int_0^t e^{-s^2}\, ds\,, \tag{B6}$$

and we note that the error function is odd, as can be seen from the brief proof below

$$\mathrm{erf}(-t) = \frac{2}{\sqrt{\pi}} \int_0^{-t} e^{-s^2}\, ds = \frac{2}{\sqrt{\pi}} \int_0^t e^{-(-s)^2}\, d(-s) = -\frac{2}{\sqrt{\pi}} \int_0^t e^{-s^2}\, ds = -\mathrm{erf}(t)\,. \tag{B7}$$

We choose to make use of the imaginary error function, defined by

$$\mathrm{erfi}(t) \equiv -i\,\mathrm{erf}(it)\,. \tag{B8}$$

Therefore $K_{\pm}$ becomes

$$K_{\pm} = -i\sigma\sqrt{\frac{\pi}{2}}\, e^{-\frac{1}{2}\sigma^2} \left[ \mathrm{erfi}\left( \frac{-\frac{i\pi}{2} \mp \sigma^2}{\sigma\sqrt{2}} \right) + \mathrm{erfi}\left( \frac{\mp \sigma}{\sqrt{2}} \right) \right]. \tag{B9}$$

By combing $K_+$ and $K_-$ as in Equation (B1) and exploiting the antisymmetry of the imaginary error function for $K_+$, we obtain the final form for $\mathbb{Z}(\sigma)$:

$$Z(\sigma) = \frac{\sigma}{2}\sqrt{\frac{\pi}{2}} \; e^{-\frac{1}{2}\sigma^2} \left[ 2\,\mathrm{erfi}\left(\frac{\sigma}{\sqrt{2}}\right) - \mathrm{erfi}\left(\frac{\sigma}{\sqrt{2}} - \frac{i\pi}{2\sigma\sqrt{2}}\right) - \mathrm{erfi}\left(\frac{\sigma}{\sqrt{2}} + \frac{i\pi}{2\sigma\sqrt{2}}\right) \right]. \qquad \textbf{(B10)}$$

## APPENDIX C: STATISTICAL EQUIVALENCE IN PARTICLES' INITIAL $z$-COORDINATE SELECTION

The purpose of this section is to provide a conceptual understanding for why individual particles are chosen to only be initialized at one lattice point above the maximum height instead of randomly the height coordinate as well. To begin, suppose a particle $A$ is initialized with a position in the $xy$-plane, given by $\boldsymbol{r}_A$. This particle then is set to have an initial height $h_A$ which is one lattice unit above the maximum height of the evolving film, as shown in Fig. 8. We require that this particle is above the surface maximum so that it has the potential to travel to any $xy$-coordinate for deposition, given a suitable trajectory, whose polar coordinate is $\theta$. As can be seen in the figure, this trajectory, and more importantly the polar angle particle $A$ has with respect to the surface normal, is identical to that of particle $A'$ - the latter of which has position in the $xy$-plane of $\boldsymbol{r}_{A'}$ and initial height given by $h_{A'}$. Let $\Delta h \equiv h_{A'} - h_A$ and $\Delta \boldsymbol{r} \equiv \boldsymbol{r}_{A'} - \boldsymbol{r}_A$. Thus, in order for the two depositions to be equivalent with respect to their deposition angle $\theta$, then we arrive at the condition that

$$\tan\theta = \frac{\Delta h}{\|\Delta \boldsymbol{r}\|}. \qquad \textbf{(C1)}$$

Since $h_{A'}$ is arbitrary provided it is greater than $h_A$, and there is no probabilistic preference given to $\boldsymbol{r}_A$ over $\boldsymbol{r}_{A'}$, it follows that for any conceivable initial position $\boldsymbol{r}_{A'} + h_{A'}\hat{z}$, the surface will develop equivalently from a particle instead initialized at position $\boldsymbol{r}_A + h_A\hat{z}$. And therefore we need only initialize a particle one lattice step above the maximum height.

## APPENDIX D: A SCHEMATIC OF THE MONTE CARLO SIMULATION ALGORITHM USED

The diagram of the Monte Carlo simulation we employed which is described in the methods section is shown in Fig. 9 to provide a visual representation of how our code models thin film evolution.

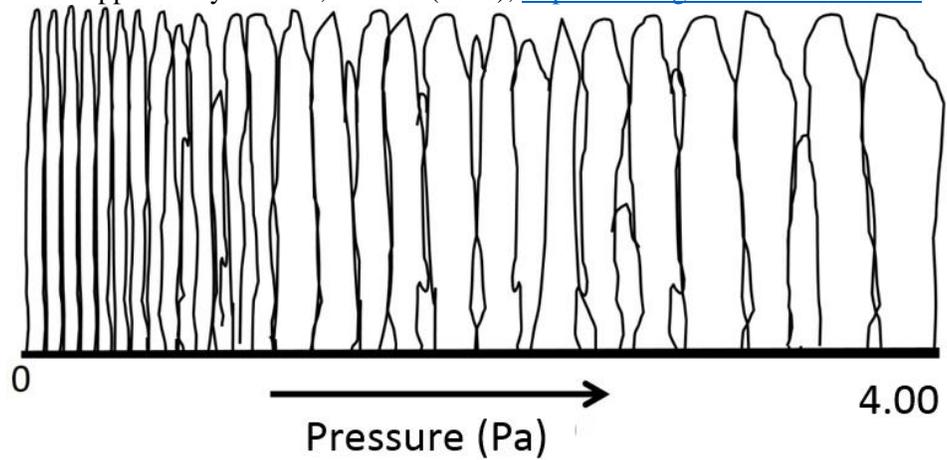

FIG. 1: A sketch of the columnar structure of films subject to different sputtering gas pressures for fixed temperature and film height.

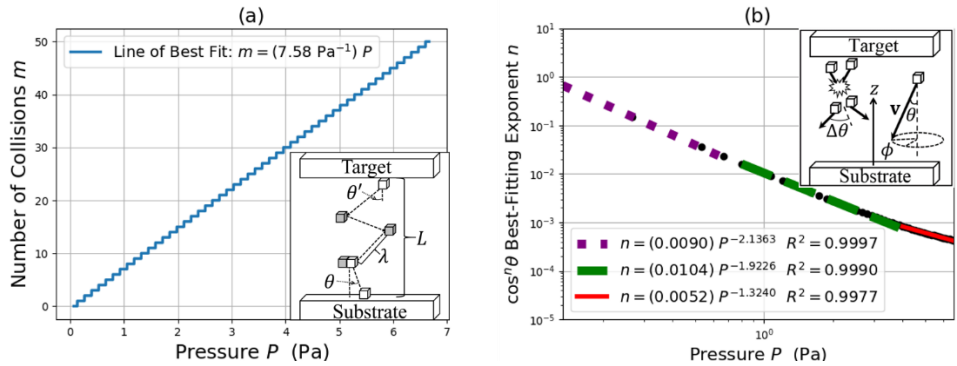

FIG. 2: (a) A plot of the number of collisions of incident specie will undergo before depositing on the evolving film as a function of chamber pressure, given $L = 10$ cm, $T = 300$ K, and $d = 2.12$ Å. The steps in the line results from the floor function given in Equation (4). The inset sketch shows a particle emitted from the target at angle $\theta'$ as it travels to the substrate a distance $L$ away. As it travels it collides with the sputtering gas (shaded) after traversing a mean free path length $\lambda$ until it is deposited on the surface with an angle $\theta$. (b) A plot of the best-fitting exponent from the cosine-power distribution as a function of chamber pressure, as numerically calculated from Equation (14). Three least-squares regression curves are plotted purely for predictive purposes and their equations are shown in the legend. Their respective $R^2$ values are calculated from the line of best fit on log-log axes. The inset schematic shows the pictorial definitions of $\phi$, $\theta$, and $\Delta\theta$ using in Section IIA.

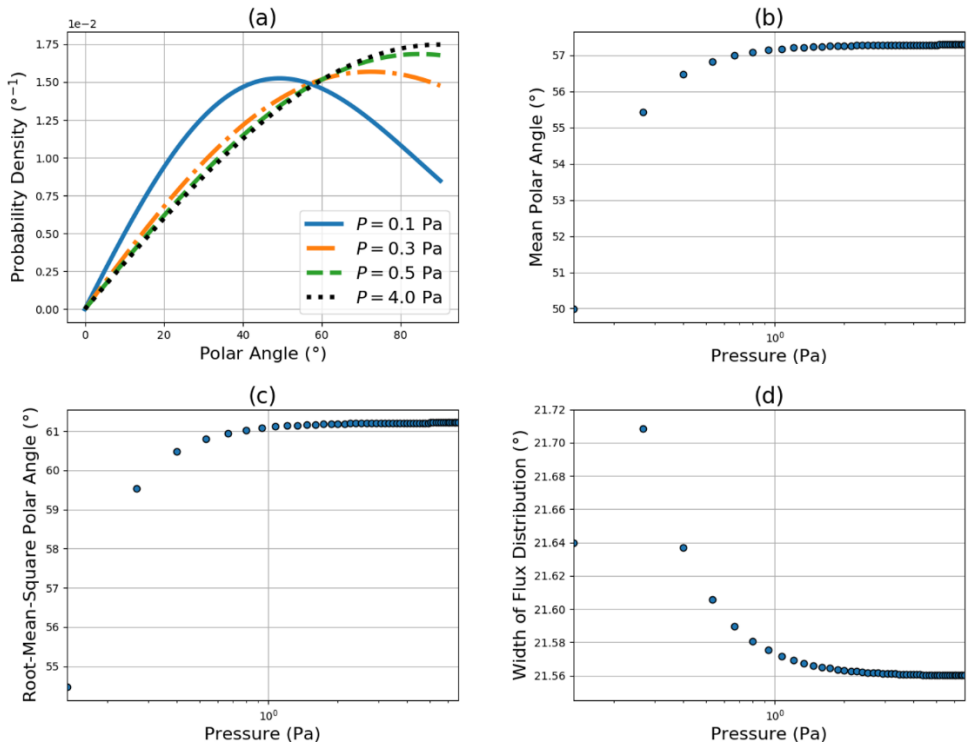

FIG. 3: (a) A few Gaussian flux distributions calculated using Equation (8) given different chamber pressures listed in the plot legend. (b) The mean deposition angle of the Gaussian flux distribution as a function of chamber pressure. (c) The root-mean-square deposition angle of the Gaussian flux distribution as a function of chamber pressure. (d) The width of the Gaussian flux distribution as a function of chamber pressure.

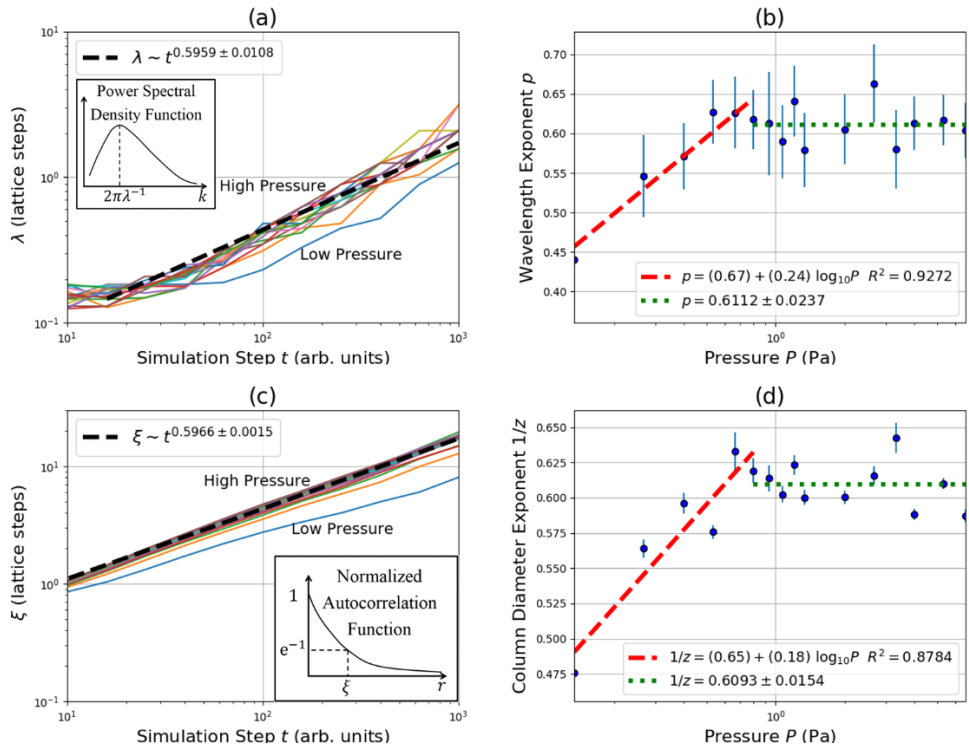

FIG. 4: (a) A plot of the film wavelength as a function of simulated time. The lower curves correspond to lower pressure and the higher curves correspond to higher pressure. The dotted line represents the average slope from all the curves. The inset curve is a sketch of how $\lambda$ was measured from the power spectral density function, where $k$ is the magnitude of the cylindrically symmetric wave-vector. (b) The value of the wavelength exponent $p$ as a function of chamber pressure. Two lines of best fit are shown, one having nonzero slope and the other representing the average value of $p$ in that domain. (c) A plot of the lateral correlation length as a function of simulated time. The lower curves are again generated from lower pressures and the higher curves are from higher pressures. The dotted line again shows the average slope of all the curves. The inset figure is a sketch of how $\xi$ was measured from the normalized autocorrelation function, where $r$ is the distance from the origin in the $xy$-plane. (d) The lateral correlation length exponent $1/z$ as a function of chamber pressure. Two lines of best fit are again shown, where the latter is the average value over that domain.

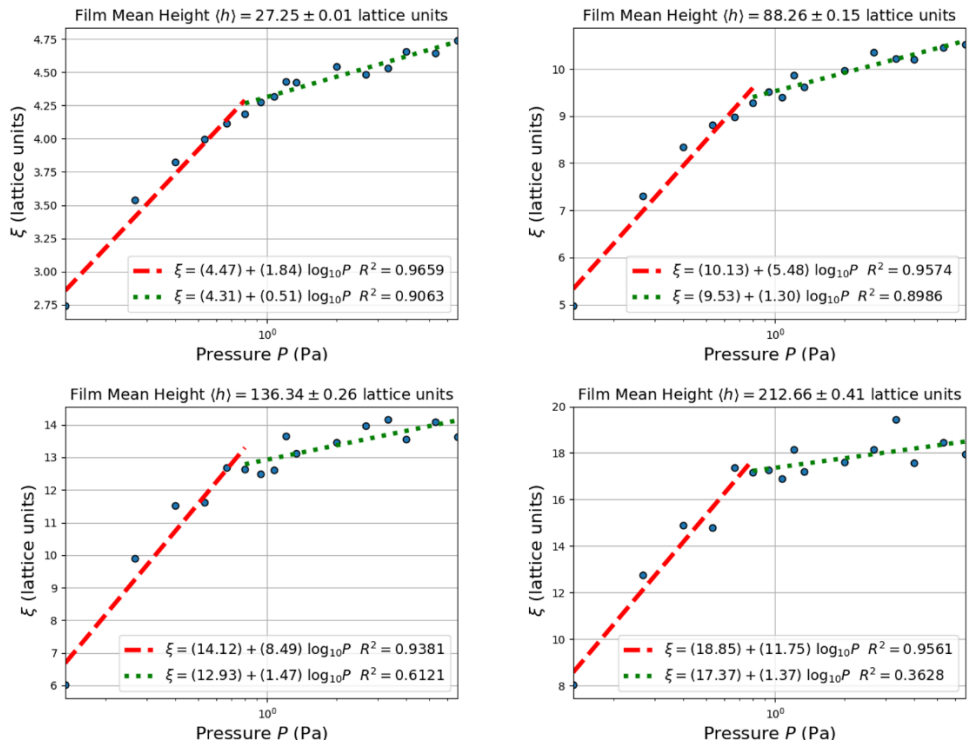

FIG. 5: A few plots of the simulated films' lateral correlation length as a function of pressure at specific heights listed at the top of each plot. The uncertainty in $\langle h \rangle$ comes from the average of all the mean film heights at their specific chamber pressures. On each plot, a line of best fit is plotted over two domains, and each equation is listed in each of the plot's respective legends.

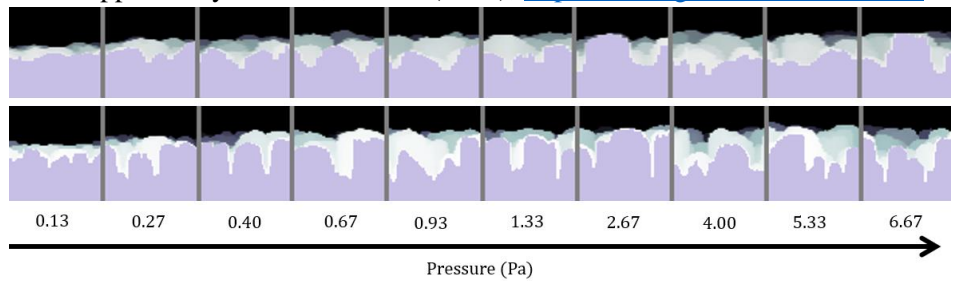

FIG. 6: Images of a section of the evolving film as viewed from the side as a function of pressure at the specific film mean height of $\langle h \rangle = 27.25 \pm 0.01$ lattice units (top band) and $\langle h \rangle = 88.26 \pm 0.15$ lattice units (bottom band). The bottom band is scaled by a factor of 0.5 for easier comparison with the top band. These heights correspond to the film growth after 100 and 251 simulation steps, respectively.

$P = 0.13$ Pa    $P = 4.00$ Pa

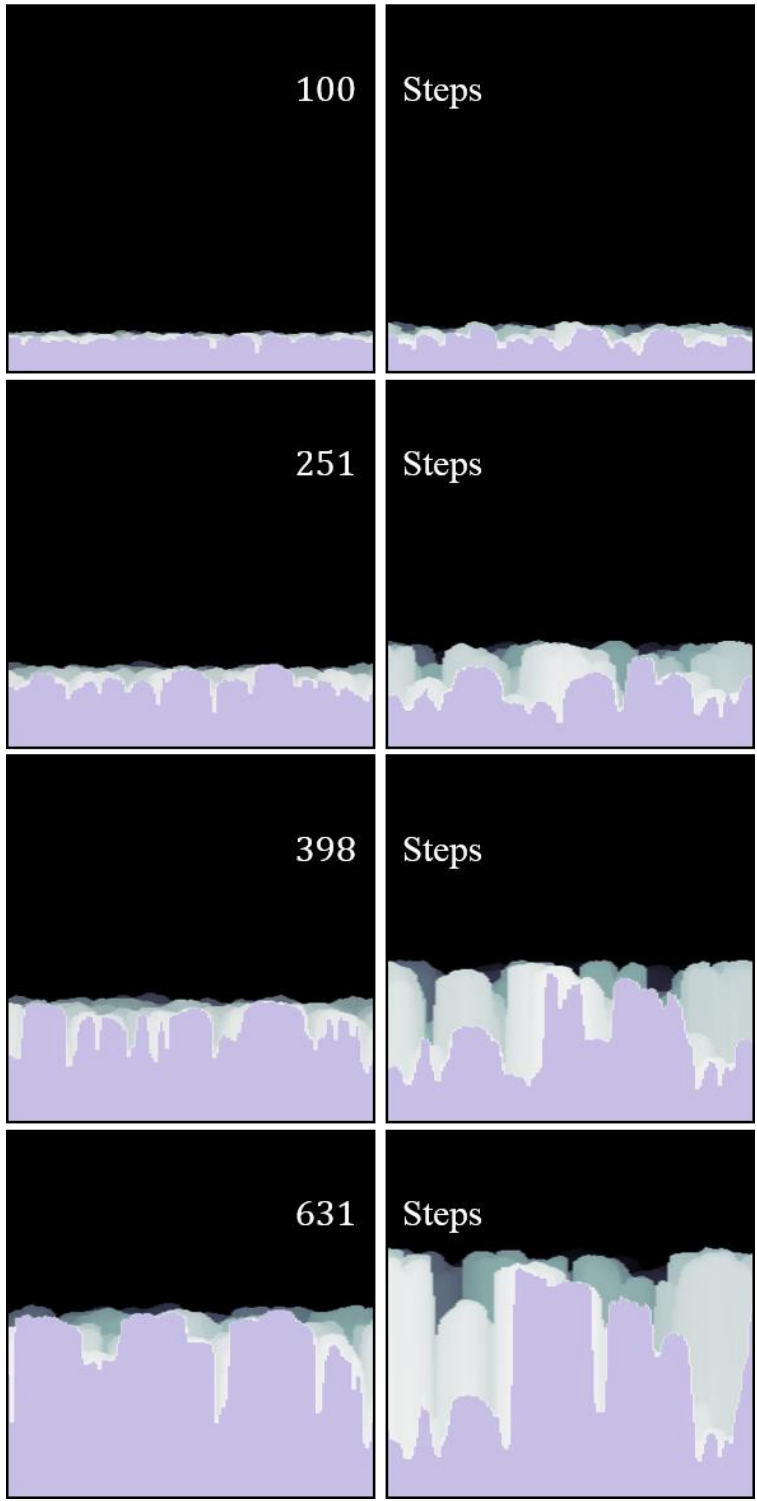

FIG. 7: Plots of the simulated film as a function of time in simulation steps. The leftward column shows a film under 0.13 Pa, while the rightward plot is under 4.00 Pa.

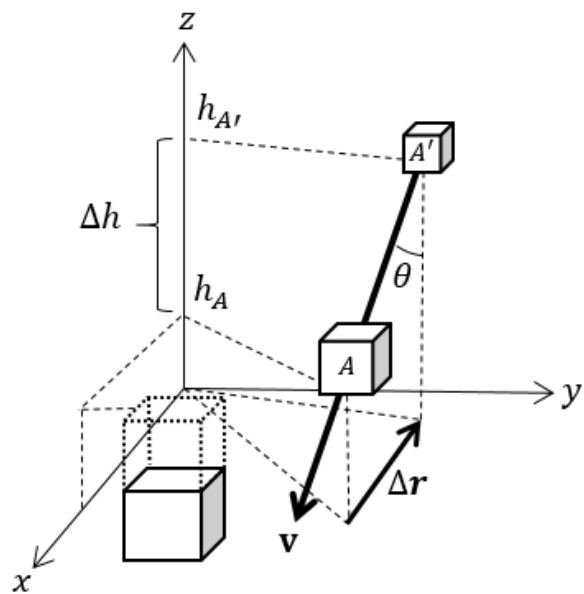

FIG. 8: A sketch of how a particle $A$ with constant velocity **v** has an identical trajectory to a particle $A'$ with an equivalent velocity, but is translated spatially by the vector $\Delta\boldsymbol{r} + \Delta h\hat{z}$.

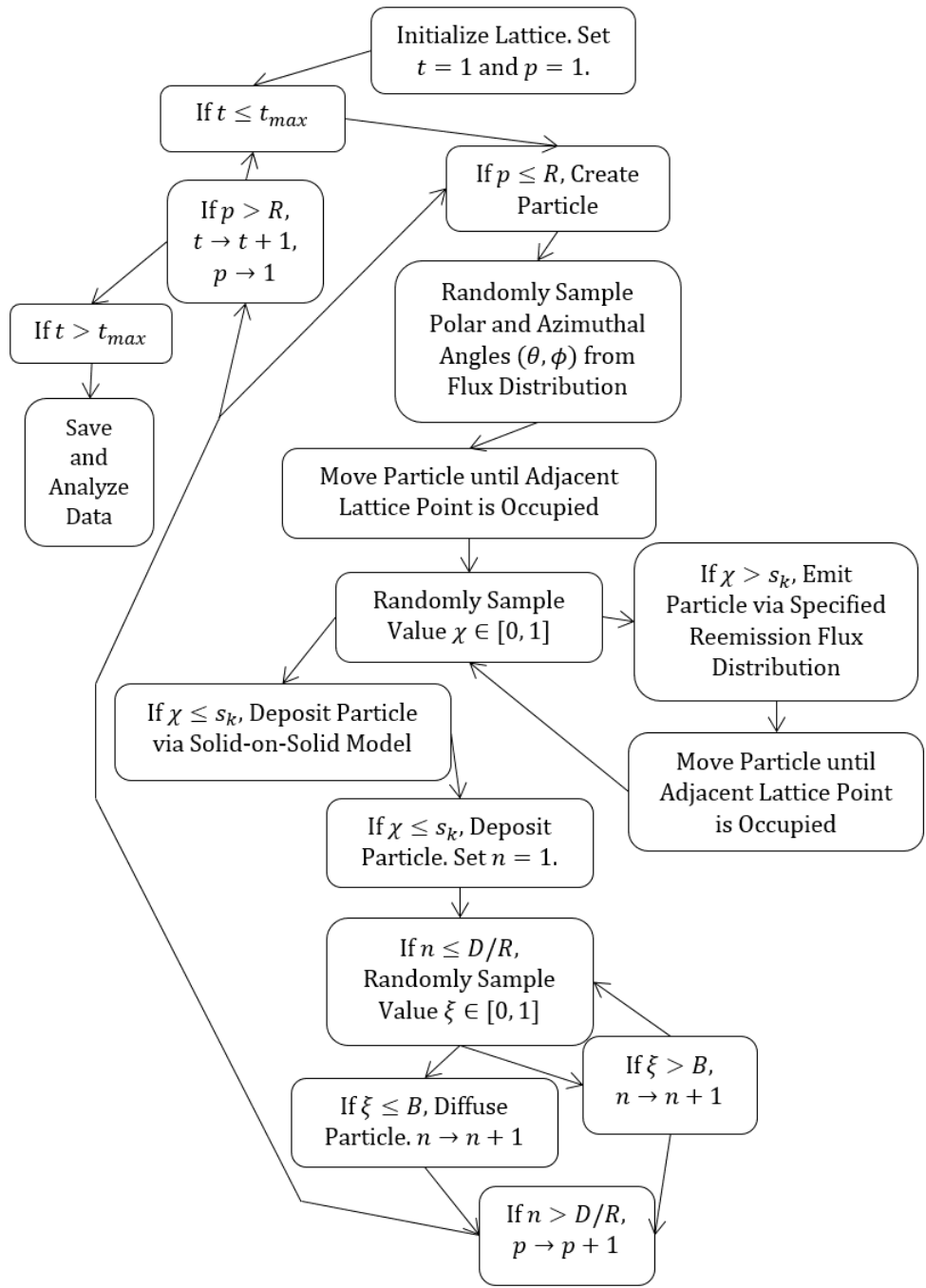

FIG. 9: A flow chart representing our Monte Carlo simulation algorithm.